\documentclass[journal, twocolumn,letterpaper]{IEEEtran}
\usepackage{ifpdf}
\usepackage{cite}
\usepackage{color}
\usepackage{amsthm}
\usepackage{amsmath,amssymb}
\usepackage{algorithmic}
\usepackage{array}
\usepackage{mdwmath}
\usepackage{mdwtab}
\usepackage{graphicx}
\usepackage{lipsum}
\usepackage{epstopdf }
\usepackage{algorithm,algorithmic}
\usepackage{enumerate}
\usepackage{eqparbox}
\usepackage{url}
\usepackage{subfig}
\usepackage{amsbsy}

\usepackage{geometry}
\def\cJ{\mathcal{J}}

\def\cN{\mathcal{N}}

\def\cT{\mathcal{T}}

\usepackage[only,tensf,]{rawfonts}

\newcommand{\tx}[1]{\text{tx}(l)}
\newcommand{\rx}[1]{\text{rx}(l)}

\newcommand{\separator}{
  \begin{center}
    \rule{\columnwidth}{0.3mm}
  \end{center}
}

\newcommand{\bi}{\begin{itemize}}
\newcommand{\ei}{\end{itemize}}
 \newtheorem{proposition}{Proposition}

\newcommand{\be}{\begin{equation}}
\newcommand{\ee}{\end{equation}}
\newcommand{\bea}{\begin{equation*}}
\newcommand{\eea}{\end{equation*}}
\newcommand{\beq}{\begin{eqnarray}}
\newcommand{\eeq}{\end{eqnarray}}
\newcommand{\beqa}{\begin{eqnarray*}}
\newcommand{\eeqa}{\end{eqnarray*}}

\def\log{\mbox{log}}

\renewcommand\floatsep{0mm}
\renewcommand\textfloatsep{2mm}

\IEEEoverridecommandlockouts 
\begin{document}
\title{ An Online Secretary Framework for Fog Network Formation with Minimal Latency\vspace{-2mm}}
\author{\IEEEauthorblockN{Gilsoo~Lee$^{\dag}$,~Walid~Saad$^{\dag}$, and~Mehdi~Bennis$^\ddag$
}\\
\IEEEauthorblockA{
\small $^{\dag}$ Wireless@VT, Department of Electrical and Computer Engineering, Virginia Tech, Blacksburg, VA, USA, \\ 
Emails: \protect\url{{gilsoolee, walids}@vt.edu}. \vspace{-2mm}\\
\small $^\ddag$ Centre for Wireless Communications, University of Oulu, Finland, Email: \url{bennis@ee.oulu.fi}.
\vspace{-12mm}
}
\thanks{This research been supported by the U.S. National Science Foundation under Grant CNS-1460333 and the Academy of Finland CARMA project.}
}

\maketitle

\begin{abstract} 
Fog computing is seen as a promising approach to perform distributed, low-latency computation for supporting Internet of Things applications. However,  due to the unpredictable arrival of available neighboring fog nodes, the  dynamic formation of a fog network can be challenging.  In essence, a given fog node must smartly select the set of neighboring fog nodes that can provide low-latency computations.  In this paper, this  problem of fog network formation and  task distribution  is studied  considering a hybrid cloud-fog architecture.  The goal of the proposed framework is to minimize the maximum computational latency by enabling a given fog node to form a suitable fog network,  under  uncertainty on the arrival process of neighboring fog nodes.    To solve this problem, a novel approach based on the online secretary framework  is proposed. 
To find the desired set of neighboring fog nodes,  an online algorithm  is developed to enable a task initiating fog node  to decide on which other  nodes can be used as part of its fog network, to offload computational tasks, without knowing any prior information on the future arrivals  of those other nodes. Simulation results show that the proposed online algorithm can successfully select an optimal set of neighboring fog nodes while achieving a latency that is as small as the one resulting  from an ideal, offline scheme that has complete knowledge of the system. The results also show how, using the proposed approach, the computational tasks can be properly distributed between the fog network and a remote cloud server. 
\end{abstract}
\IEEEpeerreviewmaketitle

\vspace{-8mm}
\section{Introduction}
The Internet of Things (IoT) is expected to connect over 50 billion things worldwide, by 2020 \cite{cisco2015fog}. To handle such massive and diverse data traffic, there is a need for distributed computation which can be effectively handled using the so-called fog computing paradigm~\cite{cisco2015fog}. Fog computing allows overcoming the limitations of centralized cloud computation, by enabling distributed,  low-latency computation at the network edge, for supporting IoT applications. The advantages of the fog architecture comes from the transfer of the network functions to the network edge. Indeed, significant amounts of data can be stored, controlled, and computed over  the fog networks that are configured and managed by end-user nodes \cite{chiang2016fog}. However, to reap the benefits of fog networks many~architectural and operational challenges must be addressed~\cite{vallati2015exploiting, khaledi2016profitable, souza2016handling, park2016joint, yu2016joint,deng2015towards, mao2016power}. 

To configure a fog network, the authors in \cite{vallati2015exploiting} propose the use of a device-to-device (D2D)-based network  that can efficiently support networking between a fog node and sensors. When tasks must be computed in a distributed way, there is a need for resource sharing between fog nodes. For instance, the work in \cite{khaledi2016profitable} proposes a task allocation approach that minimizes  the overall task completion time by  using a multidimensional auction. Moreover, the authors in \cite{souza2016handling} study the delay minimization problem in multilayer scenario with both fog and cloud, in which each layer's node has a different delay. Also, the authors in \cite{park2016joint} investigate the problem of minimizing the aggregate cloud fronthaul and wireless transmission latency. 
In \cite{yu2016joint}, a task scheduling algorithm is proposed to jointly optimize the radio and computing resources with the goal of reducing the energy consumption of users while satisfying the delay constraint. The problem of optimizing power consumption~is~also~considered in \cite{deng2015towards} subject to the delay constraint using a queueing-theoretic delay model  at the cloud. Moreover, the work in \cite{mao2016power} studies the power consumption minimization~problem in an online scenario for which future arrivals of tasks is  uncertain. 

In all of these existing task distribution fog works \cite{khaledi2016profitable, souza2016handling, park2016joint, yu2016joint,deng2015towards}, it is generally assumed that information on the formation of the fog network is completely known. However, in practice,~the fog network can be spontaneously initiated by a fog node when other neighboring fog nodes start to dynamically join and leave the network. Hence, the presence of a neighboring fog node can be uncertain. 
Indeed, it is challenging for a fog node  to know when and where another fog node will arrive. Thus, there exists an inherent uncertainty stemming from the unknown locations  and availability of fog nodes. Further, most of the existing works \cite{park2016joint, yu2016joint,souza2016handling } typically assume a simple transmission or computational latency model for a fog node. In contrast, the use of a queueing-theoretic model for both transmission and computational latency is necessary to capture the realistic latency. Consequently, unlike the existing literature \cite{khaledi2016profitable, souza2016handling, park2016joint, yu2016joint,deng2015towards} which assumes full information knowledge for fog network formation and rely on simple delay models, our goal is to design an \emph{online approach} to enable an on-the-fly information of the fog network, under uncertainty,~while minimizing~computational~latency,~given~a~realistic~delay~model. 

The main contribution of this paper is   a novel framework for online fog network formation and task distribution in a hybrid fog-cloud network. This framework allows any given fog node to dynamically  construct a fog network by selecting the most suitable set of neighboring fog nodes in the presence of uncertainty on the arrival order of neighboring fog nodes. This fog node can jointly use its fog network as well as a distant cloud server to compute a number of tasks. We formulate an online optimization problem whose objective is to minimize the maximum computational latency of all fog nodes  by properly selecting the set of fog nodes to which computations will be offloaded while also properly distributing the tasks among those fog nodes and the cloud. To solve this problem without any prior information on the future arrivals of fog nodes and their performance, we propose a new approach based on the exploration and exploitation structures from the \emph{online k-secretary framework} \cite{kleinberg2005multiple}. By using the algorithm, a given fog node  can observe the unknown environment in the exploration stage. Then, in the exploitation stage, the fog node can determine how to offload its computational tasks between other, local fog nodes and a cloud server. Simulation results show that the proposed online algorithm can minimize the maximum latency by suitably distributing tasks across fog nodes and a cloud server while achieving a performance that is near-optimal compared to an offline solution that has full information on all neighboring fog node arrivals. 

The rest of this paper is organized as follows. In Section~\ref{sec:systemmodel}, the system model is presented.  
In Section~\ref{sec:problemformulation}, we formulate the proposed online problem. Section~\ref{sec:algorithm} presents our proposed online solution. Simulation results are analyzed in Section~\ref{sec:numericalresults} while conclusions are drawn in Section \ref{sec:conclusion}.   

\vspace{-3mm}
\section{System Model}\label{sec:systemmodel}
\vspace{-2mm}

\begin{figure}[]\vspace{-2mm}
\centering 
\includegraphics[width=0.28\textwidth]{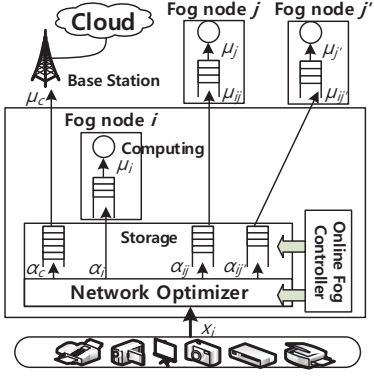}\vspace{-2mm}
\caption{\small System model of the fog networking architecture with cloud.}
\label{fig:system}\vspace{-2mm}
\end{figure}

Consider a fog network   consisting of a sensor layer, a fog layer, and a cloud layer as shown in Fig.~\ref{fig:system}. 
In this system, the sensor layer includes smart and  small-sized IoT sensors that do not have enough computational capability. Thus, these sensors offload their task data to the fog and cloud layers for  remote distributed computing purposes. We assume that the various kinds of sensors send their task data to a fog node $i$, and the size of this data will be  $x_i$ packets per second. Here, fog node $i$ assumes the roles of collecting, storing, controlling, and processing the task data from the sensor layer, as is typical in practical fog networking scenarios \cite{chiang2016fog}. 

In our architecture, fog node $i$ must  cooperate with other neighboring fog nodes and the cloud data center. It is assumed that there is a set $\mathcal{N}$ of $N$ fog nodes. For a given fog node $i$, we focus on the fog computing case in which fog node $i$ builds a network with a set $\mathcal{J} \subset \mathcal{N}$  of $J$ neighboring fog nodes. Also, since the cloud is typically located at a remote location, fog node $i$ must access the cloud via wireless communication links using a cellular base station $c$.

When fog node $i$ receives  $x_i$ tasks,  each node on a fog-cloud network will locally compute a fraction of $x_i$ that is received from the initial fog node $i$. The fraction of tasks locally computed by fog node $i$ will be given by $ \lambda_i = \alpha_i x_i$. Then, the number of tasks  offloaded from fog node $i$ to fog node $j\in \cJ$ will be $\lambda_{ij} = \alpha_{ij} x_i$. Therefore, the number of tasks processed at the fog layer will be  $\lambda_f = (\alpha_i + \sum_{j\in\cJ}\alpha_{ij}) x_i$. The number of remaining tasks that are  offloaded to the cloud will be $\lambda_c = \alpha_c x_i$. When fog node $i$ makes a decision on the distribution of all input tasks $x_i$, the task distribution variables can be represented as vector ${\boldsymbol{\alpha}}=[\alpha_i, \alpha_c, \alpha_{i1}, \!\cdots\!, \alpha_{ij}, \!\cdots\!, \alpha_{iJ}]$ with  \!$\sum_{j\in \cJ}\! \alpha_{ij} + \alpha_i +\alpha_c \!\!=\!\! 1$ \!{\color{black} where $\alpha_i, \alpha_{ij}, \alpha_c \!\in\! [0,1]$}. Naturally, the total number of tasks that arrive at fog node $i$ is equal to  the number of tasks assigned to computation nodes in the fog and cloud layers. Since $x_i$ is the sum of packets from various  sensors, it is  assumed that $x_i$ follows a Poisson arrival process \cite{deng2015towards}. When the tasks are distributed according to $\alpha_i$, $\alpha_c$, and $\alpha_{ij}$, $j\in\cJ$, the tasks offloaded to each node $\lambda_i$, $\lambda_c$, and $\lambda_{ij}, j\!\in\!\cJ,$ will also follow a Poisson process while the task are scheduled in a round robin fashion \cite{bertsekas1992data}. 

When the tasks arrive from the sensors to fog node $i$, they are first saved in fog node $i$'s storage. Thus, there is a waiting delay before tasks are transmitted and distributed to another node. The delays related to the transmission from $i$ to $c$ or $j$ can be modeled using a \emph{transmission queue}. Moreover, when the tasks arrive at the destination, the latency required to perform the actual computations  will be captured by a \emph{computation queue}. In Fig.~\ref{fig:system}, we show examples of both type of queues. For instance, for transmission queues, fog node $i$  has transmission queues for each fog node $j$ and the cloud $c$. For computation, each fog node has a computation queue. To model the transmission queue, we assume that tasks are transmitted to fog node $j$ over a wireless channel. Then, if a  task has a size of $K$ bits, the service rate can be defined by\vspace{-2mm}
\beq\label{eq:mu}\vspace{-5mm}
\mu_{ij} =\frac{1}{K} B \log_2\left(1+\frac{g_{ij}P_{tx,i}}{B N_0}\right),
\eeq
\vspace{-4mm}\\
where $g_{ij}=\beta_1 d_{ij}^{-\beta_2}$ is the channel gain between fog~nodes~\!$i$\! and \!$j$\! with $d_{ij}$ being the distance between them. $\beta_1$~and~$\beta_2$ are, respectively, the path loss exponent and path loss constant. $P_{tx,i}$ is the transmission power of fog node $i$, \!$B$\!~is~the~bandwidth of the channel, and \!$N_0$\! is the noise power spectral  density. Since the tasks arrive according to a Poisson process, and the transmission time in \!\eqref{eq:mu}\! is deterministic, the latency of the transmission queue can be modeled as~an~\!M/D/1\!~system~\cite{bertsekas1992data}:\vspace{-3mm}
\beq
T_j(\alpha_{ij}) = \frac{\lambda_{ij}}{2\mu_{ij}(\mu_{ij}-\lambda_{ij})} + \frac{1}{\mu_{ij}},
\eeq
\vspace{-4mm}\\
where the first term is the waiting time in the queue at fog node $i$, and  the second term is the transmission delay between fog nodes $i$ and $j$. Similarly, when  the tasks are offloaded to the cloud, the transmission queue delay will be:\vspace{-2mm}
\beq
T_c(\alpha_c) = \frac{\lambda_{c}}{2\mu_{c}(\mu_{c}-\lambda_{c})} + \frac{1}{\mu_c},  
\eeq 
\vspace{-4mm}\\
where the service rate $\mu_c$ between fog node $i$ and cloud $c$ is given by \eqref{eq:mu} where fog node $j$ is replaced with cloud~$c$. 

Next,  we define the computation queue. When a fog node needs to compute a task, this task will experience a waiting time in the computation queue of this fog node due to a previous task that is being currently processed. Since a  fog node $j$  receives tasks from not only fog node $i$ but also other fog nodes and sensors, the task arrival process can be reasonably approximated by a Poisson process by applying the Kleinrock approximation \cite{bertsekas1992data}. Therefore,  the computation queue can be modeled as an M/D/1 and the corresponding latency of the  fog node $j$'s computation  can be given by
\vspace{-2.5mm}
\beq
S_j(\alpha_{ij}) = \frac{\lambda_{ij}}{2\mu_{j}(\mu_{j}-\lambda_{ij})} + \frac{1}{\mu_{j}} + d_j, 
\eeq
\vspace{-3mm}\\
where the first term is the waiting delay in the computation queue, and the second term is  the delay for fetching the proper application that is needed to compute the task. The delay of this fetching procedure depends on the performance of the node's hardware which is a deterministic constant that determines the service time of the computation queue. In the first and second terms, $\mu_j$ is a  parameter related to the overall hardware performance of  fog node $j$. $d_j = c_j \lambda_{ij}$ is  the actual computation time of the task where $c_j$ is a constant time incurred to compute a task. For example, $1/c_j$ can be  proportional to the CPU clock frequency of fog node $j$. Then, when fog node $i$ locally computes its assigned tasks $\lambda_i$, the latency can will be: \vspace{-3mm}
\beq
S_i(\alpha_i) = \frac{\lambda_{i}}{2\mu_{i}(\mu_{i}-\lambda_{i})} + \frac{1}{\mu_{i}} + d_i,
\eeq
\vspace{-4mm}\\
where $\mu_i$ is the hardware performance of fog node $i$ and $d_i=c_i \lambda_i$ is  fog node $i$'s computing time. \!To model the~computation time at the cloud, since  the cloud has superior hardware performance compared to the fog node's hardware, the waiting time at the computation queue can be ignored. This implies that the cloud initiates the computation for the received tasks without having queueing delay; thus, we only account for the actual computing delay. Thus, when tasks are computed by the cloud, the computing delay at the cloud  can be defined by\vspace{-2mm}
\beq
S_c(\alpha_c) = d_c,
\eeq
\vspace{-7mm}\\
where $d_c = c_c \lambda_c$. 

In essence, if a task is routed to cloud $c$,  the latency will~be\vspace{-2mm}
\beq
 D_c(\alpha_c) =  T_c(\alpha_c) + S_c(\alpha_c).
\eeq\vspace{-5mm}\\
Also, if a task is offloaded to fog node $j$, then the latency can be presented by the sum of the transmission and computation queueing delays:\vspace{-2mm}
\beq\label{Dj}
D_j(\alpha_{ij}) = T_j(\alpha_{ij}) + S_j(\alpha_{ij}).
\eeq
\vspace{-5mm}\\
Furthermore, when fog node $i$ computes the tasks locally, the latency is given by\vspace{-3mm}
\beq
D_i(\alpha_i) = S_i(\alpha_i),
\eeq
\vspace{-6mm}\\
since no transmission queue is  necessary for  local computing. 

\vspace{-3mm}
\section{Problem Formulation}\label{sec:problemformulation}
\vspace{-1mm}
Given the defined system model, our goal is to form a fog network and  to effectively distribute tasks. To form a fog network and offload its tasks,  a fog node $i$ must opportunistically find neighboring fog nodes. In practice, such neighbors will dynamically arrive and leave the system. As a result, the initial fog node $i$ will be unable to know a priori whether an adjacent fog node will be available to assist with its computation. Moreover, since the total number of neighboring fog nodes as well as their locations are unknown and highly unpredictable, optimizing the fog network formation  and task distribution processes  becomes a challenging problem. Under such uncertainty, selecting neighboring fog nodes must also account for potential arrival of new fog nodes that can potentially provide a higher data rate and stronger computational capabilities. To cope with the uncertainty of the neighboring fog node arrivals while considering the data rate and computing capability of current and future fog nodes, we introduce an \emph{online optimization scheme} that can handle the problem of fog network formation and task distribution under uncertainty. 

First, we formulate the following online fog network formation and task distribution problem whose goal is to minimize the maximum  latency when computing a task that arrives at fog node~$i$: \vspace{-2mm}
\beq\label{problem1}
\min_{ \cJ, \boldsymbol{\alpha}}   &
                        \max \left(D_i(\alpha_i), \; D_c(\alpha_c), \; D_{j \in \cJ}(\alpha_{ij})\right) + \eta (J+1),\\
\textrm{s.t.} & \alpha_i +\alpha_c  +\sum_{j\in\cJ}  \alpha_{ij} =1, \label{problem:const}\\
                   & \alpha_i  \in [0,1], \alpha_c \in [0,1], \\
                   & \alpha_{ij}  \in [0,1], \forall j \in \cJ \subset \cN, 
\eeq \vspace{-6mm}\\
where  $\eta$ is the time cost for creating and managing the transmission queues for the various neighboring fog nodes and the cloud. For example, when fog node $i$ manages one queue for the cloud and $J$ queues for the fog nodes, $\eta (J+1)$ will capture  the additional time cost at fog node $i$. In essence, in problem~\eqref{problem1}, the objective function is the sum of the maximum latency among different computation nodes and the time cost that increases with the number of nodes in the fog network. We determine the set of neighboring fog nodes $\cJ$ when they arrive online and the task distribution vector $\boldsymbol\alpha$ so that the computing latency is minimized.  

In \eqref{problem1}, while the maximum number of neighboring fog nodes can be pre-determined by fog node $i$, we assume that fog nodes arrive in an online and arbitrary manner. This implies that the information about each fog node is collected sequentially. For example, a smartphone can choose to become a fog node spontaneously if it wants to share its resources. Such case shows how the initial fog node $i$ that manages the fog network and distributes tasks is  unable to know any information on future fog nodes. Therefore, in our problem, the arrival order can be represented by an index $n \in \cN$. At each arrival event, the arrival order $n$ increases by one; thus, index  $n$ can be seen as the time order of arrival. When fog node $n$ arrives, we  know the information of only fog node $n$.   

In our model, whenever fog node $n$ appears in the network, fog node $i$ must decide whether to select $n$  or not. If fog node $n$ is chosen, then it is indexed by $j$ and included in the set $\cJ$ which is a subset of $\cN$. Otherwise, fog node $i$ will no longer be able to select fog node $n$ since the latter can join  another fog network or  terminate its resource sharing offer to fog node $i$. Under such incomplete information, finding the optimal solution of \eqref{problem1} is challenging and, as such, one has to seek an online, sub-optimal solution that is robust to uncertainty. Next, we develop an online algorithm to solve \eqref{problem1} and optimize the fog network formation and task distribution problems. 

\vspace{-3mm}
\section{Online Secretary Problem for \\Fog Network Formation}\label{sec:algorithm}
\vspace{-2mm}

To solve \eqref{problem1}, we need to find the set of neighboring fog nodes $\cJ$ and the task distribution vector $\boldsymbol{\alpha}$ that minimize the maximum latency. The decision about  $\cJ$ faces two primary challenges: how many fog nodes are required in the fog network and which fog nodes join the fog network. Finding the optimal $\cJ$ in an online scenario can be challenging, so we relax the complexity of the problem by fixing the maximum number of neighboring fog nodes. Fog node $i$ can at most support a certain number of neighbors due to various resource limiations, e.g., limited memory or storage size. Then, our online algorithm can make a decision on which fog nodes are chosen in $\cJ$. Also, if set $\cJ$ is determined, optimizing the task distribution vector $\boldsymbol{\alpha}$ becomes an offline optimization; thus, the problem can be minimized by using an effective optimization method such as the interior-point algorithm. 

We can first observe that  the first term in the objective function \eqref{problem1} decreases as the number of neighboring fog nodes  increases since  distributed computing can reduce  latency. However, the value of \eqref{problem1} can increase if the wireless latency increases. Also, \eqref{problem1} can increase if the number of fog nodes becomes too large. For instance, the time cost required to manage the fog networking can limit the number of fog computing nodes. Thus, there is a tradeoff between the  latency of distributed computation  and the time cost of managing multiple queues when using more number of neighboring fog nodes. By considering this tradeoff, we assume that  a practical size of  distributed computing networks is predetermined and given as   parameter $J$ in our algorithm. 

We can first observe a property when the number of neighboring fog nodes is given. \vspace{-2mm}
\begin{proposition} 
For a given $\cJ$, if there exists  $\boldsymbol{\alpha}$ such that  $D=D_i(\alpha_i) = D_c(\alpha_c) = D_j(\alpha_{ij})$, $\forall j \in \cJ$ where $D$ is a constant, task distribution $\boldsymbol\alpha$ is the optimal solution of problem~\eqref{problem1}. 
\end{proposition}\vspace{-4mm}
\begin{proof}
Let call  $\boldsymbol{\alpha}$ as the initial distribution, and assume that any other task distribution  $\boldsymbol{\alpha'}$ different from  $\boldsymbol{\alpha}$ is the optimal distribution. When $\boldsymbol{\alpha'}$ is considered, we can find a certain node denoted by A satisfying $\alpha'_A < \alpha_A$ where $\alpha'_A \in \boldsymbol{\alpha'}$ and $\alpha_A \in \boldsymbol{\alpha}$. This then yields $D_A(\alpha'_A) < D_A(\alpha_A)$. Due to the constraint~\eqref{problem:const}, there exists another node $B$ such that $B\neq A$, $\alpha'_B > \alpha_B$, and $D_B(\alpha'_B) > D_B(\alpha_B)$ where $\alpha'_B \in \boldsymbol{\alpha'}$ and $\alpha_B \in \boldsymbol{\alpha}$. Since $D_B(\alpha'_B) > D_B(\alpha_B) =  D_A(\alpha_A) > D_A(\alpha'_A)$, we must decrease $\alpha'_B$ to minimize the maximum, i.e., $D_B(\alpha'_B)$. Hence, we can clearly see that  $\boldsymbol{\alpha'}$ is not optimal, and, thus, initial distribution $\boldsymbol{\alpha}$ is optimal. 
\end{proof}\vspace{-2mm}
\hspace{-1mm}Since the optimal task distribution results in an equal latency at different nodes, if the maximum number of neighboring fog nodes $J$ is determined, the problem can be reduced to choosing the neighboring fog node $j$ that can~minimize~latency~$D_j$. 

\begin{algorithm}[t]\smallskip
\caption{\small Online Fog Network Formation Algorithm}
\begin{algorithmic}[1]\label{algorithm}\vspace{-1mm}
\footnotesize\item[1 :] \hspace{0.0cm}  Input: $\tau$, $J$, and $\mu_i$.
\item[2 :] \hspace{0.0cm}  Measure $\mu_c$. 
\item[    ] \hspace{0.0cm}  {\bf Exploration}
\item[3 :] \hspace{0.0cm}  while $|\cT| < \tau$ 
\item[4 :] \hspace{0.3cm}    Wait arrival of fog node $n$.
\item[5 :] \hspace{0.3cm}    Measure $\mu_{in}$ and $\mu_{n}$.
\item[6 :] \hspace{0.3cm}    $\cT  \leftarrow  \cT \cup \{\mu_{in}+\mu_{n}\}.$
\item[7 :] \hspace{0.0cm}  end while
\item[    ] \hspace{0.0cm}  {\bf Exploitation}
\item[8 :] \hspace{0.0cm}  while $|\cJ| < J$
\item[9 :] \hspace{0.3cm}    Wait arrival of fog node $n$.
\item[10:]\hspace{0.3cm}     Measure $\mu_{in}$ and $\mu_n$. Find $t^* = \max \cT $.
\item[11:]\hspace{0.3cm}     if $\mu_{ij}+\mu_j > t^*$
\item[12:]\hspace{0.6cm}       $\cJ  \leftarrow  \cJ + \{n\}$.
\item[13:]\hspace{0.6cm}       $\cT \leftarrow \cT \setminus t^*$.
\item[14:]\hspace{0.6cm}       Solve \eqref{problem3} to find distribution $\boldsymbol{\alpha}$
\item[15:]\hspace{0.3cm}     end if
\item[16:]\hspace{0.0cm}   end while
\end{algorithmic}\vspace{-1mm}
\end{algorithm}

Due to the fact that $D_j$ in \eqref{Dj}  can decrease when $\mu_{ij}$~and~$\mu_j$ increase,\! the problem of selecting the best fog nodes can then be written as:\vspace{-4mm}
\beq\label{problem3}
\max_{\cJ} && \sum_{j\in \cJ}\left(  \mu_{ij} + \mu_{j} \right).
\eeq
\vspace{-4mm}\\
This problem implies that our proposed solution must select  the $J$ fog nodes whose data rate and computational  capability are larger than those of the  $N-J$ fog nodes when the information about the neighboring fog nodes are known to fog node $i$  in an online way. To find $\cJ$, we propose an online algorithm that builds on the so-called $k$-secretary problem that is introduced in \cite{kleinberg2005multiple}. In this problem, when there are $k$  job positions, a company interviews $N$ candidates sequentially in a random order. Right after finishing the interview, the company has to make a decision whether to accept the candidate or not, given that the company will not be able to recall a candidate later once this candidate has been rejected. Clearly, there is a direct analogy between our problem and the secretary~problem as we seek to find $J$ neighboring fog devices which corresponds to filling $k$ job positions. Therefore,~by modifying this online secretary framework, we propose Algorithm~1 to find $\cJ$ and $\boldsymbol{\alpha}$. 

Algorithm~\ref{algorithm} sequentially optimizes the network formation problem by determining $\cJ$ and minimize the latency by determining the task distribution vector $\boldsymbol{\alpha}$ when the size of fog networking is given by $J$. The parameter $J$ can be determined by trial and error. For example, we can set an upper and lower bounds of $J$ and use a bisection method to choose $J$ such that the total cost is close to  optimal and the latency is minimized. Then, Algorithm~1  learns the uncertain environment of the online setting and determines $\cJ$ during \emph{exploration} and \emph{exploitation} stages, respectively. Once the fog network is determined, the distribution $\boldsymbol{\alpha}$ of the tasks can be found in an offline manner using the interior point method.

In Algorithm~\ref{algorithm}, we need parameter $\tau$  that indicates the number of observations needed to learn the environment. First, we observe $\tau$ fog nodes that arrive sequentially, using which it is possible to build an observation set $\cT$ that consists of the observed values of $\mu_{in}+\mu_{n}$. This observation procedure is called  the exploration stage, and it provides  the thresholds that can be used to make a decision in the subsequent exploitation stage. Therefore, through the exploration stage,  we can have information on the uncertain neighboring fog nodes. 

After constructing set $\cT$ with  $\tau$ samples, we make a decision in an online manner during the exploitation stage. 
When fog node $n$ arrives online, we can know  $\mu_{in}$ and $\mu_n$. Then, we can compare this  information about $n$ to the largest sample in set $\cT$. If the arriving fog node's performance is better than the largest sample $t^*$, then we immediately include fog node $n$ in $\cJ$. When a new fog node joins the network, the task distribution problem for a given $\cJ$ is an offline problem, so  $\boldsymbol{\alpha}$ can be optimized by using a solver. By repeating this procedure and updating $\cJ$, the set of neighboring fog nodes can be  determined. Consequently, the proposed algorithm will find a set $\cJ$ having high $\mu_{ij}$ and $\mu_j$; thus, Algorithm~1 ends by allowing fog node $i$ to form a latency-minimal fog network and distribute the tasks across fog and cloud layers.

\vspace{-3.5mm}
\section{Simulation Results}\label{sec:numericalresults}\vspace{-1.5mm}

\hspace{-1mm}For our simulations, we consider an  initial fog node that can connect to  neighboring fog nodes that are uniformly distributed within a 50\,\text{m}$\times$50\,\text{m} square area. The arrival sequence of the neighboring fog nodes follows a uniform distribution. Each fog node can use  a subcarrier of  bandwidth  $15$~kHz. The power spectral density of the  noise is -174~dBm/Hz and $P_{tx,i}=20$~dBm. The channel gain is set to  $\beta_1 \!=\!10^{-3}$ and $\beta_2\!=\!4$ with a channel gain of $-30$~dB at the reference distance of 1\,m. The packet size $K$ is set to $1500$~bytes. The distance between fog node $i$ and the base station used to connect to the cloud is  $600~\text{m}$. All statistical results are averaged over a large number of simulation runs.  We assume   equal computation resources for fog nodes such that, i.e.,  $\mu_i\!=\!\mu_j\!=\!8$ packets per second, $\forall j\in\cJ$, and we set $\tau\!=\!3$, $c_i\!=\!c_j\!=\!0.05$ and $c_c\!=\!0.025$. For comparison, we use the offline, optimal algorithm that has complete knowledge~of~the~system.  

\begin{figure}[]
\centering \vspace{-0mm}
\includegraphics[width=0.34\textwidth]{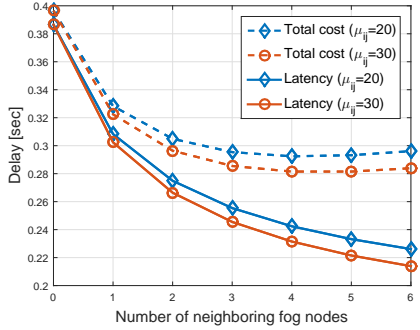}\vspace{-2mm}
\caption{\small The value of the objective function for different data rate of  neighboring fog nodes in an offline setting.}
\label{fig:delay}\vspace{-1.5mm}
\end{figure}

\begin{figure}[]
\centering \vspace{-0.0mm}
\includegraphics[width=0.34\textwidth]{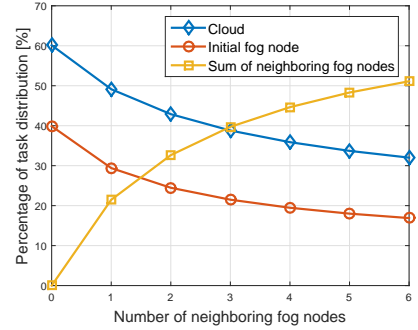}\vspace{-2mm}
\caption{\small The task distribution for fog node $i$, cloud $c$, and neighboring fog nodes  in an offline setting.}
\label{fig:distribution}\vspace{-2mm}
\end{figure}

In Fig.~\ref{fig:delay}, we show the total cost defined in \eqref{problem1} and the latency for  different numbers of fog nodes  with $\mu_c=8.8$ and $\mu_{ij}=20$ or $30$, $\forall j\in\cJ$. The simulation results in Fig.~\ref{fig:delay} are carried out in an offline setting, and we exploit this observation to determine a possible parameter $J$ to run Algorithm~1. 
For instance, we can first see that the computational latency decreases when the number of neighboring fog nodes increases. 
At the same time, Fig.~\ref{fig:delay} shows that the total cost is minimized by four neighboring fog nodes. From these observations, if the cost of a certain number of neighboring fog nodes is  similar to the minimum, we may choose a greater number of neighboring fog nodes to minimize latency. In that sense, the range between 4 and 6 neighbor fog devices cloud be a potential value of $J$. Note that the gap between the total cost and latency characterizes the time cost required to manage the fog network which naturally increases with the size of the network. Also,  we observe that the total cost and latency  decrease if the data rate of a fog node $\mu_{ij}$ increases. For instance, if $\mu_{ij}$ increases from $20$ to $30$, then the total cost is reduced to $3.7\%$ for a network with $4$ neighboring fog nodes. 

Fig.~\ref{fig:distribution} shows  the task distribution for different numbers of fog nodes   with the same parameters  used  in Fig.~\ref{fig:delay} with $\mu_{ij}=20$. From Fig.~\ref{fig:distribution}, we can see that,  when the number of neighboring fog nodes increases, the number of tasks computed by the fog layer increases, and the number of tasks offloaded to the cloud decreases. For instance, the percentage of tasks computed by the cloud is $60\%$ when there is no neighboring fog node, but it can decrease down to $32\%$ if six fog nodes  join the fog computing. 

\begin{figure}[]
\centering 
\includegraphics[width=0.35\textwidth]{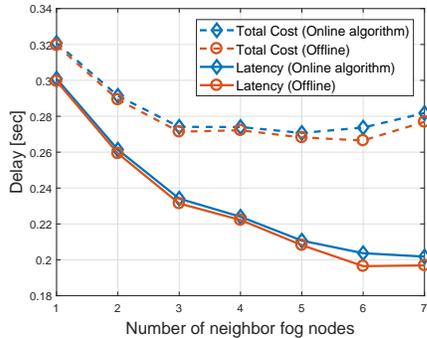}\vspace{-2.5mm}
\caption{\small The total cost achieved by using Algorithm~\ref{algorithm} compared with the optimal, offline solutions for different numbers of fog nodes. }
\label{fig:delayoff}\vspace{-1mm}
\end{figure}

Fig.~\ref{fig:delayoff} shows the total cost and latency for different numbers of neighboring fog nodes with $\tau=3$ when the proposed, online Algorithm~\ref{algorithm} is used. For a given $J$, we compare the performance of $\cJ$ found by the proposed algorithm to the performance of the optimal set of neighboring fog nodes found in the offline case. We first see that  the results of the total cost and latency in online and offline scenarios are very close. For example, it can be observed that the total cost~\eqref{problem1} can be minimized with around 6 neighboring fog nodes. In this case, the gap between the online and offline solutions, in terms of total cost,  is roughly $2.7$\%. A similar small gap is also seen for the latency. This demonstrates the effectiveness of the proposed algorithm under the online scenario. Also, Figs.~\ref{fig:delayoff} shows that, due to the time cost for queue management,  the total cost  increases when the number of neighboring fog nodes increases from 6 to 7 while the latency is still decreasing. 

\begin{figure}[]
\centering \vspace{-0mm}
\includegraphics[width=0.35\textwidth]{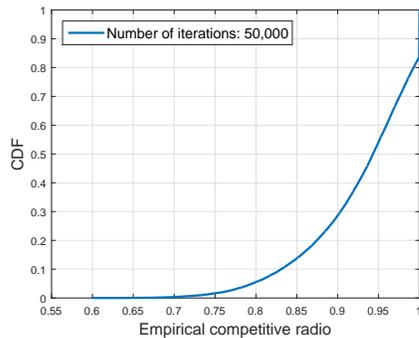}\vspace{-2.5mm}
\caption{\small The CDF of the empirical competitive ratio for the online solution of problem \eqref{problem3}.}
\label{fig:cr}\vspace{-3mm}
\end{figure}

Fig.~\ref{fig:cr} shows  the empirical competitive ratio for problem~\eqref{problem3}. The competitive ratio is defined as the ratio of the cost from the algorithm used in an online setting and the optimal cost found in the offline setting where the cost is defined by \eqref{problem3}. Thus, a competitive ratio can measure how close the proposed algorithm can achieve the solution compared to the offline solution. We can first see that $16.2\%$ of iterations  achieve a competitive ratio of $1$ which  means that the result of online algorithm coincides with the offline optimal solution of problem~\eqref{problem3}. Fig.~\ref{fig:cr} also shows that in 50\% of the iterations, the online algorithm can achieve at least $94.2$\% of the optimal value of \eqref{problem3}. Finally, over $50,000$ iterations, the empirical competitive ratio in the worst case is shown to be $0.59$. Thus, the results from Fig.~\ref{fig:cr} shows that Algorithm~1 can effectively form a fog network, in an online manner, while minimizing latency and costs.     \vspace{-0.5mm}

\begin{figure}[]
\centering \vspace{-0mm}
\includegraphics[width=0.35\textwidth]{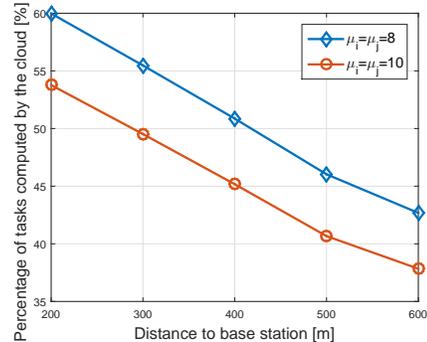}\vspace{-2.5mm}
\caption{\small The average percentage of the cloud's tasks for  different distances between $i$ and $c$, in an online setting. }
\label{fig:distmu}\vspace{-3mm}
\end{figure}

In Fig.~\ref{fig:distmu}, we show the percentage of tasks computed by the cloud for different distances from $200$~m to $600$~m using Algorithm~1 with $J\!\!=\!\!2$. The result shows that the number of tasks computed at the cloud decreases as the distance increases. This is due to the fact that a longer distance decreases $\mu_c$, thus yielding an increasing of the computation delay of the cloud. For example, Fig.~\ref{fig:distmu} shows that increasing the distance from 200~m to 600~m  can result in  $28.8\%$ fewer tasks that are offloaded to the cloud for  $\mu_i=8$. Also, we can see that fewer tasks are offloaded to the cloud when the fog nodes are equipped with better computational capabilities. For example, if $\mu_i$ or $\mu_j$ increases from $8$ to $10$, then the tasks at the cloud can decrease by up to $11.3$\%. 

\vspace{-3.5mm}
\section{Conclusion}\label{sec:conclusion}\vspace{-2mm}
In this paper, we have proposed a novel framework to optimize the formation of fog networks and distribution of computational tasks in a hybrid fog-cloud system. We have formulated the problem as an online secretary problem  which enables the neighboring fog node to join the fog networking  effectively in the presence of uncertainty about fog node arrivals. We have shown that by using the online algorithm, the neighboring fog nodes are suitably selected without knowing any prior information on future fog node arrivals. Simulation results have shown that the proposed online algorithm achieves a near-optimal latency while effectively offloading computational tasks across fog and cloud layers. 

\vspace{-4mm}
\bibliographystyle{IEEEtran}

\begin{thebibliography}{10}
\vspace{-1mm}
\providecommand{\url}[1]{#1}
\csname url@samestyle\endcsname
\providecommand{\newblock}{\relax}
\providecommand{\bibinfo}[2]{#2}
\providecommand{\BIBentrySTDinterwordspacing}{\spaceskip=0pt\relax}
\providecommand{\BIBentryALTinterwordstretchfactor}{4}
\providecommand{\BIBentryALTinterwordspacing}{\spaceskip=\fontdimen2\font plus
\BIBentryALTinterwordstretchfactor\fontdimen3\font minus
  \fontdimen4\font\relax}
\providecommand{\BIBforeignlanguage}[2]{{%
\expandafter\ifx\csname l@#1\endcsname\relax
\typeout{** WARNING: IEEEtran.bst: No hyphenation pattern has been}%
\typeout{** loaded for the language `#1'. Using the pattern for}%
\typeout{** the default language instead.}%
\else
\language=\csname l@#1\endcsname
\fi
#2}}
\providecommand{\BIBdecl}{\relax}
\BIBdecl

\bibitem{cisco2015fog}
{Cisco}, ``Fog computing and the {Internet of Things}: Extend the cloud to
  where the things are,'' \emph{Cisco white paper}, 2015.

\bibitem{chiang2016fog}
M.~Chiang and T.~Zhang, ``{Fog} and {IoT}: An overview of research
  opportunities,'' \emph{IEEE Internet of Things Journal}, vol.~PP, no.~99, pp.
  1--1, June 2016.

\bibitem{vallati2015exploiting}
C.~Vallati, A.~Virdis, E.~Mingozzi, and G.~Stea, ``Exploiting {LTE D2D}
  communications in {M2M} fog platforms: Deployment and practical issues,'' in
  \emph{Proc. IEEE 2nd World Forum on IoT}, Milan, Italy, Dec. 2015, pp.
  585--590.

\bibitem{khaledi2016profitable}
M.~Khaledi, M.~Khaledi, and S.~K. Kasera, ``Profitable task allocation in
  mobile cloud computing,'' in \emph{Proc. 12th Int. Symposium on QoS and
  Security for Wireless and Mobile Networks}, Malta, Nov. 2016.

\bibitem{souza2016handling}
V.~B.~C. Souza, W.~Ramírez, X.~Masip-Bruin, E.~Marín-Tordera, G.~Ren, and
  G.~Tashakor, ``Handling service allocation in combined fog-cloud scenarios,''
  in \emph{Proc. IEEE Int. Conf. on Commun. (ICC)}, Kuala Lumpur, Malaysia, May
  2016, pp. 1--5.

\bibitem{park2016joint}
S.-H. Park, O.~Simeone, and S.~Shamai, ``Joint cloud and edge processing for
  latency minimization in fog radio access networks,'' in \emph{Proc. IEEE 17th
  Int. Wksh. on Signal Process. Adv. in Wireless Commun.}, Edinburgh, UK, July
  2016, pp. 1--5.

\bibitem{yu2016joint}
Y.~Yu, J.~Zhang, and K.~B. Letaief, ``Joint subcarrier and {CPU} time
  allocation for mobile edge computing,'' in \emph{Proc. of IEEE Global Commun.
  Conf. (GLOBECOM)}, Washington DC, USA, Dec. 2016.

\bibitem{deng2015towards}
R.~Deng, R.~Lu, C.~Lai, and T.~H. Luan, ``Towards power consumption-delay
  tradeoff by workload allocation in cloud-fog computing,'' in \emph{Proc. IEEE
  Int. Conf. on Commun. (ICC)}, London, UK, June 2015.

\bibitem{mao2016power}
Y.~Mao, J.~Zhang, S.~Song, and K.~B. Letaief, ``Power-delay tradeoff in
  multi-user mobile-edge computing systems,'' in \emph{Proc. of IEEE Global
  Commun. Conf. (GLOBECOM)}, Washington DC, USA, Dec. 2016.

\bibitem{kleinberg2005multiple}
R.~Kleinberg, ``A multiple-choice secretary algorithm with applications to
  online auctions,'' in \emph{Proc. the 16th Symposium on Discrete Algorithms
  (SODA)}, Vancouver, Canada, Jan. 2005, pp. 630--631.

\bibitem{bertsekas1992data}
D.~P. Bertsekas, R.~G. Gallager, and P.~Humblet, \emph{Data networks}.\hskip
  1em plus 0.5em minus 0.4em\relax Prentice-Hall International New Jersey,
  1992, vol.~2.

\end{thebibliography}

\end{document}